\documentclass[english,twocolumn]{svjour3}
\usepackage[T1]{fontenc}
\usepackage[latin9]{inputenc}
\usepackage{url}
\usepackage{amsmath}
\usepackage{amssymb}
\usepackage{graphicx}
\usepackage{esint}

\makeatletter
\usepackage[sort&compress,numbers]{natbib}

\usepackage{ifpdf}
\ifpdf

\IfFileExists{lmodern.sty}
 {\usepackage{lmodern}}{}

\fi 

\makeatother

\usepackage{babel}
\begin{document}

\title{Interference Minimization in 5G Heterogeneous Networks}

\author{Tao Han \and Guoqiang Mao \and Qiang Li \and Lijun Wang \and
Jing Zhang}

\date{Accepted by Mobile Networks and Applications. The final publication
is available at Springer via http://dx.doi.org/10.1007/s11036-014-0564-1.}

\institute{Tao Han \and Qiang Li (Corresponding author) \and Jing Zhang \at
School of Electronic Information and Communications, Huazhong University
of Science and Technology, China\\
\email{\{hantao, qli\_patrick, zhangjing\}@hust.edu.cn} \and Guoqiang
Mao \at School of Computing and Communications, University of Technology
Sydney, Australia\\
National ICT Australia (NICTA), Australia\\
\email{g.mao@ieee.org} \and Lijun Wang \at Department of Information
Science and Technology, Wenhua College, China\\
\email{wanglj22@163.com}}
\maketitle
\begin{abstract}
In this paper, we focus on one of the representative 5G network scenarios,
namely multi-tier heterogeneous cellular networks. User association
is investigated in order to reduce the down-link co-channel interference.
Firstly, in order to analyze the multi-tier heterogeneous cellular
networks where the base stations in different tiers usually adopt
different transmission powers, we propose a Transmission Power Normalization
Model (TPNM), which is able to convert a multi-tier cellular network
into a single-tier network, such that all base stations have the same
normalized transmission power. Then using TPNM, the signal and interference
received at any point in the complex multi-tier environment can be
analyzed by considering the same point in the equivalent single-tier
cellular network model, thus significantly simplifying the analysis.
On this basis, we propose a new user association scheme in heterogeneous
cellular networks, where the base station that leads to the smallest
interference to other co-channel mobile stations is chosen from a
set of candidate base stations that satisfy the quality-of-service
(QoS) constraint for an intended mobile station. Numerical results
show that the proposed user association scheme is able to significantly
reduce the down-link interference compared with existing schemes while
maintaining a reasonably good QoS. 
\end{abstract}

\section{Introduction}

A heterogeneous cellular network (HCN) usually consists of multiple
tiers including a macrocell tier and probably some small cell tiers,
e.g., picocell tier, femtocell tier and so on \cite{chandrasekhar08mcomm}.
In general, there are three channel allocation strategies among tiers,
namely orthogonal deployment, co-channel deployment, and partially
shared deployment \cite{fooladivanda13twc}. In order to improve the
spectral efficiency to match the ever growing demand for high data
rate nowadays and future, co-channel deployment among tiers and spatial
frequency reuse are widely employed in HCNs. In the small cell tier,
since base stations (BSs) are often deployed in an unplanned manner,
it causes more serious co-channel interference in heterogeneous networks
than that in conventional single-tier cellular networks. In view of
the severe co-channel interference under both intra-tier and inter-tier
situations \cite{saquib12wc}, interference management is very important
in a HCN \cite{heath13tsp}.

User association, also called cell association or BS association,
is one of the important approaches to performing interference management
as well as to improving the spectral efficiency and energy efficiency
\cite{son11jac}. Fooladivanda \emph{et al.} proposed a unified static
framework to study the interplay between user association and resource
allocation in HCNs \cite{fooladivanda13twc}. Ghimire \emph{et al.}
formulated a flow-based framework for the joint optimization of resource
allocation, transmission coordination, and user association in a heterogeneous
network comprising of a macro BS and a number of pico BSs and/or relay
nodes \cite{ghimire13twc}, where the performance of different combinations
of resource allocation schemes and transmission coordination mechanisms
was characterized. Jin \emph{et al.} proposed a marginal utility based
user association algorithm to transform the combinatorial optimization
problem into a network-wide utility maximization problem \cite{jin13lcomm}.
Jo \emph{et al.} developed a tractable framework for signal-to-interference-plus-noise
ratio (SINR) analysis in downlink HCNs with flexible cell association
policies \cite{jo12twc}. Madan \emph{et al.} described new paradigms
for the design and operation of HCNs, where cell splitting, cell range
expansion, semi-static resource negotiation on third-party backhaul
connections, and fast dynamic interference management for quality-of-service
(QoS) via over-the-air signaling were investigated \cite{madan10jsac}.

In HCNs, an active mobile station (MS) needs to associate itself with
a particular cell, which belongs to one of the tiers in a multi-tier
network. Conventionally, a MS is associated with the nearest BS or
the BS that provides the highest received SINR. However, these MS
association schemes do not consider the possible co-channel interference
caused to other active MSs. Motivated by this, in this paper using
stochastic geometry methods \cite{elsawy13cst}, we propose a MS association
scheme in multi-tier networks that is able to significantly reduce
the down-link co-channel interference while guaranteeing a predefined
QoS of mobile users in HCNs with open-access small cell. Consider
the interference in up-link is not always minimized when we minimize
the interference in down-link by a user association scheme, we will
not investigate the issue on interference in up-link in this paper.
The contributions of this paper are:
\begin{enumerate}
\item A Transmission Power Normalization Model (TPNM) for analyzing the
performance of multi-tier HCNs is proposed, which significantly simplifies
the analysis of the performance of multi-tier HCNs.
\item Based on TPNM, a new user association scheme is proposed to minimize
the down-link co-channel interference, which can be used in both conventional
single-tier cellular networks and multi-tier HCNs.
\item Extensive simulations are conducted, the results demonstrate that
the proposed scheme can significantly reduce the down-link interference
under the constraint that predefined QoS requirements are satisfied.
\end{enumerate}
The rest of the paper is organized as follows. Section \ref{sec:Model-and-Assumptions}
describes the system model. Section \ref{sec:Scheme-TPNS} introduces
the proposed TPNM for performance analysis in multi-tier HCNs. Based
on TPNM, we proceed to propose a new user association scheme to minimize
the down-link co-channel interference in section \ref{sec:Interference-Minimized-User-Association}.
Section \ref{sec:Numerical-Results} shows the numerical results for
the performance of the proposed user association scheme. In section
\ref{sec:Conclusions}, we conclude the paper.

\section{System model\label{sec:Model-and-Assumptions}}

We consider a heterogeneous cellular multi-tier network that is composed
of $K$-tier networks where $K\in\mathbb{N}$ with only a single BS
located winthin each cell of the multi-tier networks. The transmission
powers at the BSs of the $k$-th tier network are assumed to be equal
and denoted as $P_{k}$. We assume that the distribution of the BSs
in the $k$-th tier network follows a homogeneous Poisson point process
$\Phi_{k}^{\text{BS}}$ with intensity $\lambda_{k}^{\text{BS}}$.
Assuming that the multiple cells of different tiers are overlaid in
the same area geographically, then the distribution of the BSs in
multi-tier HCNs is governed by a Poisson point process $\Phi^{\text{BS}}=\bigcup_{k=1}^{K}\Phi_{k}^{\text{BS}}$
with intensity $\lambda^{\text{BS}}=\sum_{k=1}^{K}\lambda_{k}^{\text{BS}}$.

Furthermore, we assume that the distribution of active MSs which are
associated with the BSs in the $k$-th tier network follows a homogeneous
Poisson process $\Phi_{k}^{\text{MS}}$ of intensity $\lambda_{k}^{\text{MS}}$.
Thus the distribution of all MSs in multi-tier HCNs is also governed
by a Poisson point process $\Phi^{\text{MS}}=\bigcup_{k=1}^{K}\Phi_{k}^{\text{MS}}$
with intensity $\lambda^{\text{MS}}=\sum_{k=1}^{K}\lambda_{k}^{\text{MS}}$. 

This paper focuses on the down-links in multi-tier HCNs, where all
BSs reuse the same frequency that is divided into orthogonal channels.
A BS allocates different orthogonal channels to the MSs in a cell.
Under such circumstances, there is no intra-cell interference. However,
due to the frequency reuse across cells, there may exist severe inter-cell
co-channel interference in multi-tier HCNs if the same sub-channel
is occupied in different cells \cite{palanisamy13icices}. For example,
given that the BSs assign the channels randomly and independently,
at a particular time instant, only a fraction of the BSs, denoted
by Poisson point process $\Phi^{\text{N\_BS}}$ of intensity $\lambda^{\text{N\_BS}}$,
are using a specific channel $C_{n}$ simultaneously to transmit to
the corresponding MSs, denoted by Poisson point process $\Phi^{\text{N\_MS}}$
of intensity $\lambda^{\text{N\_MS}}=\lambda^{\text{N\_BS}}$, where
the BSs in $\Phi^{\text{N\_BS}}$ and the MSs in $\Phi^{\text{N\_MS}}$
are communication pairs. 

Assuming BSs assign down-link channels to the MSs associated with
them randomly, then the MSs using the same channel $C_{n}$, i.e.
$\Phi^{\text{N\_MS}}$, can be considered to follow a homogeneous
Poisson point process \cite{xiang2013twc}, which is thinned from
point process $\Phi^{\text{MS}}$. Then we define $\Phi^{\text{N\_MS}}$
as an interfering set, in which MSs are interfered by the BSs that
are transmitting to other MSs in the set because they use the same
channel $C_{n}$.

For ease of exposition, only path loss effect is considered in the
wireless channel models. Without loss of generality, we consider a
given BS $x$ and a desired MS $y$. Then the desired signal power
$P_{xy}$ received at $y$ is expressed as 
\begin{equation}
P_{xy}=P_{x}\,l\left(x-y\right),
\end{equation}
 where $P_{x}$ denotes the transmission power of the BS and $l\left(\cdot\right)=\left\Vert \cdot\right\Vert ^{-\alpha}$
denotes the path loss in wireless channels where $\alpha$ is the
path loss exponent. 

In this paper, we focus on the interference-limited scenario. When
a MS $y$ is associated with a BS $x$, the signal-to-interference
ratio (SIR) at $y$ is given as 
\begin{equation}
\mathrm{SIR}\left(x,y\right)=\frac{P_{xy}}{I_{y}}=\frac{P_{x}\,l\left(x-y\right)}{\sum_{x_{i}\in\Phi^{\text{N\_BS}}\backslash\left\{ x\right\} }P_{x_{i}}\,l\left(x_{i}-y\right)},
\end{equation}
 where $I_{y}$ denotes the interference received from the BSs in
$\Phi^{\mathrm{\text{N\_BS}}}$ except $x_{i}$.

In view of the severe co-channel interference, we consider a user
association scheme where the MS $y\in\Phi^{\text{N\_MS}}$ chooses
a BS $x\in\Phi^{\text{BS}}$ to associate with, and at the same time
the interference from $x$ to other MSs $\Phi^{\text{N\_MS}}\backslash\left\{ y\right\} $
is minimized.

In order to minimize the interference caused by the chosen BS $x$
to other co-channel MSs, for ease of analysis, we consider a MS $z\in\Phi^{\text{N\_MS}}\backslash\left\{ y\right\} $
that receives the most severe interference $I_{xz}$ from BS $x$
\cite{son11jsac}. Then the minimization of the interference seen
at $z$ probably implies a minimization of the co-channel interference.

On the other hand, to satisfy a reasonable QoS constraint, it is assumed
that the distance between the specific MS $y$ and the corresponding
BS $x\in\Phi^{\text{BS}}$ should be no more than the distance between
$y$ and any BS $\forall x_{i}\in\Phi^{\text{N\_BS}}$ transmitting
in the same channel. In other words, we intend to choose a suitable
BS for $y$ such that the co-channel interference caused to other
MSs is minimized, under the constraint the QoS of the MS $y$ is satisfied.
If in $\Phi^{\text{BS}}$ there is no BS satisfies this constraint,
then MS $y$ will try to search another channel.

\section{TPNM of HCN\label{sec:Scheme-TPNS}}

\subsection{Definition of TPNM}

Different from single-tier homogeneous cellular networks where all
the BSs transmit signal using the same power, in multi-tier HCNs,
the BSs of different tiers have different transmission powers and
follow different distributions geographically.

In order to analyze the multi-tier HCN, we propose a TPNM in this
paper, which is able to convert a multi-tier HCN to a virtual single-tier
cellular network by first scaling each tier according to its corresponding
transmission power and path-loss effect, and then combining the different
tiers into a single-tier cellular network, such that all BSs have
the same normalized transmission power and the signal power and interference
received at a specific MS from the BSs in the virtual single-tier
cellular network are exactly the same as those received from the BSs
in the original multi-tier cellular network.

\begin{figure}
\begin{centering}
\includegraphics[width=0.95\columnwidth]{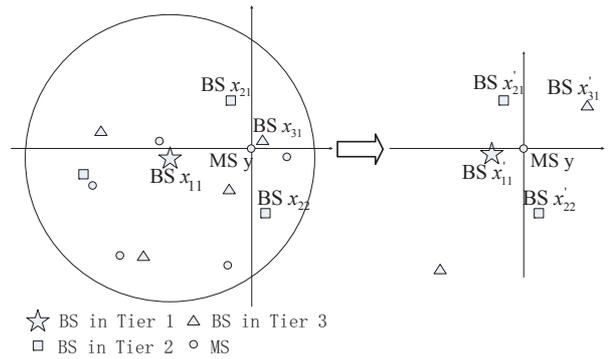}
\par\end{centering}

\caption{By TPNM, each tier is scaled by using the location of a specific MS
$y$ as the scaling center.\label{fig:TPNS} }
\end{figure}

As an example, consider a $3$-tier HCN shown in Fig. \ref{fig:TPNS}.
There are BSs in various tiers including BS $x_{11}$ in tier $1$,
BS $x_{21}$ and $x_{22}$ in tier $2$, and BS $x_{31}$ in tier
$3$, with different transmission powers. MS $y$ receives the desired
signal from the associated BS and interference from the other BSs.
For ease of analysis, we set the location of the MS $y$ as the origin
and scale each tier by using different factors such that virtual BSs
$x_{11}^{\prime}$, $x_{21}^{\prime}$, $x_{22}^{\prime}$ and $x_{31}^{\prime}$
with the same normalized power $1$ are obtained, and the received
signal/interference powers at $y$ from these virtual BSs are exactly
the same as those received from the original BSs, e.g., the power
received at MS $y$ from BS $x_{11}$ before scaling is exactly the
same as that from virtual BS $x_{11}^{\prime}$ after scaling.

In a $K$-tier HCN, the BSs in tier $k,\ k\in\left\{ 1,2,\dots,K\right\} $,
have transmission power $P_{k}$ and follow a Poisson point process
$\Phi_{k}^{\text{BS}}$ of intensity $\lambda_{k}^{\text{BS}}$. Without
loss of generality, we consider a MS $y$ located at origin $o$,
then the received signal power $P_{xy}$ at MS $y$ from a BS $x\in\Phi_{k}$
is given as 
\begin{eqnarray}
P_{xy} & = & P_{k}\,l\left(x-o\right)=P_{k}\,l\left(x\right)=1\cdot\left(P_{k}^{-\frac{1}{\alpha}}\left\Vert x\right\Vert \right)^{-\alpha}\nonumber \\
 & = & 1\cdot\left\Vert P_{k}^{-\frac{1}{\alpha}}\cdot x\right\Vert ^{-\alpha}=1\cdot l\left(P_{k}^{-\frac{1}{\alpha}}\cdot x\right),\label{eq:power_tpnm}
\end{eqnarray}
 where $1$ is the normalized transmission power and $l(x-o)$ is
the path loss function from $x$ to $y$.

From Eq. (\ref{eq:power_tpnm}), it is observed that the signal power
received at $y$ from $x$ is equal to that received from a virtual
BS $x^{\prime}=P_{k}^{-\frac{1}{\alpha}}\cdot x$ with transmission
power $1$ and located at $P_{k}^{-\frac{1}{\alpha}}\cdot x$. Following
this, the Poisson point process $\Phi_{k}^{\text{BS}}$ can be scaled
to a point process $\Phi_{k}^{\text{BS}\prime}=P_{k}^{-\frac{1}{\alpha}}\cdot\Phi_{k}^{\text{BS}}$
of intensity $\lambda_{k}^{\text{BS}\prime}=\left(\frac{1}{p_{k}^{-\frac{1}{\alpha}}}\right)^{2}\lambda_{k}^{\text{BS}}=P_{k}^{\frac{2}{\alpha}}\lambda_{k}^{\text{BS}}$,
in which all the virtual BSs have the same transmission power $1$
and produce the same received signal power (or interference power)
at the MS $y$ as the original BSs in $\Phi_{k}^{\text{BS}}$.

From the above analysis, we scale the Poisson point processes in all
$K$ tiers to normalize the BSs' transmission powers to $1$, and
then combine them into a single Poisson point process 
\begin{equation}
\Phi^{\text{BS}\prime}=\bigcup_{k=1}^{K}P_{k}^{-\frac{1}{\alpha}}\cdot\Phi_{k}^{\text{BS}},
\end{equation}
 which is of intensity $\lambda^{\text{BS}\prime}=\sum_{k=1}^{K}P_{k}^{\frac{2}{\alpha}}\lambda_{k}^{\text{BS}}$.

\subsection{Received signal power based on TPNM}

In this section, we give an example to demonstrate the advantage of
using TPNM by considering analysis of cell association where a MS
always associate with the BS delivering the highest received signal
strength. We first analyze the case without TPNM, then we present
the analysis by using TPNM.

\subsubsection{The received signal power without TPNM}

Consider a specific receiving MS $y$, without loss of generality,
we place it at the origin $o\in\mathbb{R}^{2}$. Then the BS 
\begin{equation}
x=\arg\max_{x_{i}\in\Phi^{\text{BS}}}P_{x_{i}}\left\Vert x_{i}\right\Vert ^{-\alpha},
\end{equation}
which can produce the highest received signal power at $y$, is selected
to associate with $y$.

Assuming there are $K$ tiers in the network with corresponding transmission
power $P_{k},k\in\left\{ 1,2,\dots,K\right\} $, we find the nearest
BS $x_{k}$ from each tier $\Phi_{k}^{\text{BS}}$ to $y$, i.e.,
\begin{equation}
x_{k}=\arg\min_{x_{kj}\in\Phi_{k}^{\text{BS}}}\left\Vert x_{kj}\right\Vert .
\end{equation}
According to Slivnyak theorem \cite{stoyan1995stochastic}, $\Phi_{k}^{\text{BS}}\cup\left\{ o\right\} $
has the same properties as the Poisson point process $\Phi_{k}^{\text{BS}}$,
so the distance $R_{k}\triangleq\left\Vert x_{k}-y\right\Vert =\left\Vert x_{k}-o\right\Vert =\left\Vert x_{k}\right\Vert $
between BS $x_{k}$ and MS $y$ satisfies the following probability
density function (PDF) 
\begin{equation}
f_{R_{k}}\left(r_{k}\right)=2\pi\lambda_{k}^{\text{BS}}r_{k}\cdot\mathrm{e}^{-\lambda_{k}^{\text{BS}}\pi r_{k}^{2}}.
\end{equation}

Then the signal power received at MS $y$, i.e., $P_{x_{k}y}=P_{k}R_{k}^{-\alpha}$,
has the following PDF 
\begin{align}
f_{P_{x_{k}y}\left(p_{x_{k}y}\right)} & =f_{R_{k}}\left(r_{k}\right)\cdot\left|\left(\left(\frac{P_{k}}{p_{x_{k}y}}\right)^{\frac{1}{\alpha}}\right)^{\prime}\right|\nonumber \\
 & =\frac{2\pi\lambda_{k}^{\text{BS}}}{\alpha p_{x_{k}y}}\left(\frac{P_{k}}{P_{x_{k}y}}\right)^{\frac{2}{\alpha}}\mathrm{e}^{-\pi\lambda_{k}^{\text{BS}}\left(\frac{P_{k}}{p_{x_{k}y}}\right)^{\frac{2}{\alpha}}}.
\end{align}

Defining $D_{x_{k}y}\triangleq P_{x_{k}y}^{-\frac{1}{\alpha}}=P_{k}^{-\frac{1}{\alpha}}R_{k}$,
we obtain the PDF of $D_{x_{k}y}$ as follow: 
\begin{align}
f_{D_{x_{k}y}}\left(d_{x_{k}y}\right) & =f_{P_{x_{k}y}}\left(P_{x_{k}y}\right)\cdot\left|\left(d_{x_{k}y}^{-\alpha}\right)^{\prime}\right|\nonumber \\
 & =2\pi\lambda_{k}^{\text{BS}}P_{k}^{\frac{2}{\alpha}}d_{x_{k}y}\cdot\mathrm{e}^{-\pi\lambda_{k}^{\text{BS}}P_{k}^{\frac{2}{\alpha}}d_{x_{k}y}^{2}},
\end{align}
 and the corresponding cumulative distribution function (CDF) is derived
as 
\begin{align}
F_{D_{x_{k}y}}\left(d_{x_{k}y}\right) & =\int_{-\infty}^{d_{x_{k}y}}f_{D_{x_{k}y}}\left(d_{x_{k}y}\right)\,\mathrm{d}d_{x_{k}y}\nonumber \\
 & =1-\mathrm{e}^{-\pi\lambda_{k}^{\text{BS}}P_{k}^{\frac{2}{\alpha}}d_{x_{k}y}^{2}}.
\end{align}

The BS that MS $y$ is associated with should have the largest $P_{x_{k}y}$,
so it should have the smallest $D_{x_{k}y}$ as well. Denote the smallest
$D_{x_{k}y}$ by $D_{xy}$, and the largest $P_{x_{k}y}$ by $P_{xy}$,
we obtain 
\begin{equation}
P_{xy}=\max_{k\in\left\{ 1,2,\dots,K\right\} }P_{x_{k}y},
\end{equation}
 and 
\begin{equation}
D_{xy}=\min_{k\in\left\{ 1,2,\dots,K\right\} }D_{x_{k}y}.
\end{equation}

Since random variables $D_{x_{k}y},\,k\in\left\{ 1,2,\dots,K\right\} $
are mutually independent, then the CDF of $D_{xy}$ can be derived
as 
\begin{align}
F_{D_{xy}}\left(d_{xy}\right) & =1-\prod_{k=1}^{K}\left(1-\left(1-\mathrm{e}^{-\pi\lambda_{k}^{\text{BS}}P_{k}^{\frac{2}{\alpha}}d_{x_{k}y}^{2}}\right)\right)\nonumber \\
 & =1-\mathrm{e}^{-\pi d_{x_{k}y}^{2}\sum_{k=1}^{K}\lambda_{k}^{\text{BS}}P_{k}^{\frac{2}{\alpha}}},
\end{align}
 and the PDF of $D_{xy}$ can be derived as 
\begin{equation}
f_{D_{xy}}\left(d_{xy}\right)=\left(2\pi\right)d_{xy}^{2}\sum_{k=1}^{K}\lambda_{k}^{\text{BS}}P_{k}^{\frac{2}{\alpha}}\cdot\mathrm{e}^{-\pi d_{x_{k}y}^{2}\sum_{k=1}^{K}\lambda_{k}^{\text{BS}}P_{k}^{\frac{2}{\alpha}}}.
\end{equation}

Because $D_{xy}=P_{xy}^{-\frac{1}{\alpha}}$, we have 
\begin{align}
F_{P_{xy}}\left(P_{xy}\right) & =F_{D_{xy}}\left(P_{xy}^{-\frac{1}{\alpha}}\right)\nonumber \\
 & =1-\mathrm{e}^{-\pi P_{xy}^{-\frac{2}{\alpha}}\sum_{k=1}^{K}\lambda_{k}^{\text{BS}}P_{k}^{\frac{2}{\alpha}}}.\label{eq:CDF_p_xy_org}
\end{align}

\subsubsection{The received signal power with TPNM}

By using TPNM, we scale each tier with factor $P_{k}^{-\frac{1}{\alpha}}$,
and then combine them to a virtual Poisson point process $\Phi^{\text{BS}\prime}$
of intensity $\lambda^{\text{BS}\prime}=\sum_{k=1}^{K}P_{k}^{\frac{2}{\alpha}}\lambda_{k}^{\text{BS}}$.
Denote the distance between MS $y$ and the nearest BS $x\in\Phi^{\text{BS}\prime}$
by $R$, which has the following CDF 
\begin{equation}
F_{R}\left(r\right)=1-\mathrm{e}^{-\lambda^{\text{BS}\prime}\pi r^{2}}=1-\mathrm{e}^{-\sum_{k=1}^{K}P_{k}^{\frac{2}{\alpha}}\lambda_{k}^{\text{BS}}\pi r^{2}}.
\end{equation}

Because $P_{xy}=1\cdot R^{-\alpha}$ and thus $R=P_{xy}^{-\frac{1}{\alpha}}$,
we have 
\begin{align}
F_{P_{xy}}\left(p_{xy}\right) & =1-\mathrm{e}^{-\sum_{k=1}^{K}P_{k}^{\frac{2}{\alpha}}\lambda_{k}^{\text{BS}}\pi\left(p_{xy}^{-\frac{1}{\alpha}}\right)^{2}}\nonumber \\
 & =1-\mathrm{e}^{-\pi p_{xy}^{-\frac{2}{\alpha}}\sum_{k=1}^{K}\lambda_{k}^{\text{BS}}P_{k}^{\frac{2}{\alpha}}}.\label{eq:CDF_p_xy_TPNAS}
\end{align}

Since Eq. (\ref{eq:CDF_p_xy_TPNAS}) is of the same form as Eq. (\ref{eq:CDF_p_xy_org}),
TPNM can be used to analyze the received signal and interference powers
at an arbitrary MS with path loss effect, which significantly simplifies
the derivations.

\section{Interference minimized user association scheme\label{sec:Interference-Minimized-User-Association}}

\subsection{Interference modeling of HCNs}

In a multi-tier HCN with different transmission powers across tiers,
each active MS chooses the BS that produces the highest received SINR
to associate with.

Consider an arbitrary MS $y$, by using TPNM, the multi-tier HCN can
be transformed to a virtual single-tier cellular network $\Phi^{\text{BS}\prime}=\bigcup_{k=1}^{K}P_{k}^{-\frac{1}{\alpha}}\cdot\Phi_{k}^{\text{BS}}$
of intensity $\lambda^{\text{BS}\prime}=\sum_{k=1}^{K}P_{k}^{\frac{2}{\alpha}}\lambda_{k}^{\text{BS}}$.
Then the average fraction of users that are served by tier $k$ in
open access is given as 
\begin{equation}
\bar{N}_{k}=\frac{\lambda_{k}^{\text{BS}\prime}}{\lambda^{\text{BS}\prime}}=\frac{\lambda_{k}^{\text{BS}}P_{k}^{\frac{2}{\alpha}}}{\sum_{k=1}^{K}\lambda_{k}^{\text{BS}}P_{k}^{\frac{2}{\alpha}}}.\label{eq:average_N_k}
\end{equation}

Then in tier $k$, the nearest BS $x_{k}$ is selected associate with
MS $y$. To evaluate the interference caused by $x_{k}$ to other
co-channel MSs, we consider the nearest MS to $x_{k}$ other than
$y$, i.e., 
\begin{equation}
z_{k}=\arg\min_{z_{i}\in\Phi^{\text{N\_MS}}\backslash\left\{ y\right\} }\left\Vert x_{k}-z_{i}\right\Vert ,
\end{equation}
 then the received interference at $z_{k}$ from $x_{k}$ is given
as 
\begin{equation}
I_{x_{k}z_{k}}=P_{k}\,l\left(R_{x_{k}z_{k}}\right)=P_{k}\left\Vert x_{k}-z_{k}\right\Vert ^{-\alpha},
\end{equation}
 where the distance $R_{x_{k}z_{k}}=\left\Vert x_{k}-z_{k}\right\Vert $
between $x_{k}$ and $z_{k}$ follows the following CDF and PDF respectively:
\begin{equation}
F_{R_{x_{k}z_{k}}}\left(r_{x_{k}z_{k}}\right)=1-\mathrm{e}^{-\lambda^{\text{N\_MS}}\pi r_{x_{k}z_{k}}^{2}},
\end{equation}
\begin{equation}
f_{R_{x_{k}z_{k}}}\left(r_{x_{k}z_{k}}\right)=2\pi\lambda^{\text{N\_MS}}r_{x_{k}z_{k}}\cdot\mathrm{e}^{-\pi\lambda^{\text{N\_MS}}r_{x_{k}z_{k}}^{2}}.
\end{equation}
 According to (\ref{eq:average_N_k}), we consider the probability
that the BS $x$, which is serving MS $y$, belongs to the $k$-th
tier also as $\bar{N}{}_{k}$. Then the expectation of the interference
received at MS $z$ from BS $x$ can be derived as 
\begin{align}
 & \mathbb{E}\left(I_{xz}\right)=\sum_{k=1}^{K}\bar{N}_{k}\int_{0}^{\infty}f_{R_{x_{k}z_{k}}}\left(r_{x_{k}z_{k}}\right)\cdot P_{k}\,l\left(r_{x_{k}z_{k}}\right)\,\mathrm{d}r_{x_{k}z_{k}}\nonumber \\
 & =\frac{\sum_{k=1}^{K}\lambda_{k}^{\text{BS}}P_{k}^{\frac{2}{\alpha}+1}}{\sum_{k=1}^{K}\lambda_{k}^{\text{BS}}P_{k}^{\frac{2}{\alpha}}}\cdot\int_{0}^{\infty}l\left(r_{x_{k}z_{k}}\right)f_{R_{x_{k}z_{k}}}\left(r_{x_{k}z_{k}}\right)\,\mathrm{d}r_{x_{k}z_{k}}.\label{eq:expection_I_xz_multi_tier}
\end{align}

\subsection{Interference minimized user association scheme}

In the proposed user association scheme, consider a specific MS, the
BS $x_{\text{opt}}$ that generates the largest received SIR at this
MS is selected under the constraint on the predefined QoS. 

Consider an arbitrary MS $y$, we first transform the multi-tier HCN
to a virtual single-tier cellular network $\Phi^{\text{BS}\prime}=\bigcup_{k=1}^{K}P_{k}^{-\frac{1}{\alpha}}\cdot\Phi_{k}^{\text{BS}}$
of intensity $\lambda^{\text{BS}\prime}=\sum_{k=1}^{K}P_{k}^{\frac{2}{\alpha}}\lambda_{k}^{\text{BS}}$
by TPNM. Then we have the interfering set that transmit simultaneously
in channel $C_{n}$ as $\Phi^{\text{N\_BS}\prime}\subset\Phi^{\text{BS}\prime}$
of intensity $\lambda^{\text{N\_BS}\prime}$, which is transformed
from $\Phi^{\text{N\_BS}}$ by TPNM as well.

To satisfy the QoS constraint, not all BSs in $\Phi^{\text{BS}\prime}$
can be chosen to communicate with the MS $y$, we denote the subset
of BSs that are allowed to communicate with $y$ by $T_{y}\subset\Phi^{\text{BS}\prime}$.
Then the distance between $y$ and each BS in $T_{y}$ is no more
than the distance between $y$ and any other transmitting-in-the-same-channel
BS, i.e., 

\begin{eqnarray}
T_{y} & = & \left\{ x_{i}^{\prime}:\left\Vert x_{i}^{\prime}-y\right\Vert \le\left\Vert x_{j}^{\text{\ensuremath{\prime}}}-y\right\Vert \right.\nonumber \\
 &  & \left.\qquad,x_{i}^{\prime}\in\Phi^{\text{BS}\prime},\forall x_{j}^{\prime}\in\Phi^{\text{N\_BS}\prime}\right\} .
\end{eqnarray}

We denote the number of BSs in subset $T_{y}$ by a random variable
$N_{T_{y}}$. $N_{T_{y}}$ is the number of points from $\Phi^{\text{BS}\prime}$
in the void ball $V=b\left(y,R^{\text{N\_BS}\prime}\right)$ of $\Phi^{\text{N\_BS}\prime}$,
where $R^{\text{N\_BS}\prime}$ is the void distance of $\Phi^{\text{N\_BS}\prime}$,
whose PDF is \cite{stoyan1995stochastic} 
\begin{equation}
f_{R^{\text{N\_BS}\prime}}\left(r^{\text{N\_BS}\prime}\right)=2\pi\lambda^{\text{N\_BS}\prime}r^{\text{N\_BS}\prime}\cdot\mathrm{e}^{-\pi\lambda^{\text{N\_BS}\prime}\left(r^{\text{N\_BS}\prime}\right)^{2}}.
\end{equation}

Then an estimated value $\bar{N}_{T_{y}}$ of $N_{T_{y}}$ is given
as 
\begin{align}
\bar{N}_{T_{y}} & \triangleq\mathbb{E}\left(N_{T_{y}}\right)=\lambda^{\text{BS}\prime}\cdot A\left(V\right)\nonumber \\
 & =\lambda^{\text{BS}\prime}\int_{0}^{\infty}\pi\left(r^{\text{N\_BS}\prime}\right)^{2}\cdot f_{R^{\text{N\_BS}\prime}}\left(r^{\text{N\_BS}\prime}\right)\,\mathrm{d}r^{\text{N\_BS}\prime}\nonumber \\
 & =\frac{\lambda^{\text{BS}\prime}}{\lambda^{\text{N\_BS}\prime}},\label{eq:est_val_N_T_y}
\end{align}
 where $A\left(V\right)$ denotes the area of $V$.

We assume that the proportion of the transmitting BSs in each tier
is the same. Then according to (\ref{eq:est_val_N_T_y}), we obtain
\begin{equation}
\bar{N}{}_{T_{y}}=\frac{\lambda^{\text{BS}\prime}}{\lambda^{\text{N\_BS}\prime}}=\frac{\lambda^{\text{BS}\prime}}{\lambda^{\text{N\_BS}}\cdot\frac{\lambda^{\text{BS}\prime}}{\lambda^{\text{BS}}}}=\frac{\lambda^{\text{BS}}}{\lambda^{\text{N\_BS}}}.
\end{equation}

According to (\ref{eq:average_N_k}), the average fraction of users
served by tier $k$ in open access can thus be derived as 
\begin{equation}
\bar{N}{}_{k}=\frac{\lambda_{k}^{\text{BS}}P_{k}^{\frac{2}{\alpha}}}{\sum_{k=1}^{K}\lambda_{k}^{\text{BS}}P_{k}^{\frac{2}{\alpha}}}.
\end{equation}

Then the BS that satisfies 
\begin{equation}
x_{\mathrm{opt}}=\arg\min_{x_{i}^{\prime}\in T_{y}}\max_{z_{j}^{\prime}\in\Phi^{\text{N\_MS}\prime}\backslash\left\{ y\right\} }1\cdot\left\Vert x_{i}^{\prime}-z_{j}^{\prime}\right\Vert ^{-\alpha}
\end{equation}
is selected to associate with $y$. For the co-channel MS that receives
the largest interference from $x_{opt}$, i.e., $z_{\text{opt}}$,
we have 
\begin{equation}
z_{\mathrm{opt}}=\arg\min_{z_{i}^{\prime}\in\Phi^{\text{N\_MS}\prime}\backslash\left\{ y\right\} }\left\Vert x_{\text{opt}}-z_{i}^{\prime}\right\Vert .
\end{equation}

Denote the distance between $x_{\text{opt}}$ and $z_{\mathrm{opt}}$
by $R_{\mathrm{opt}}=\left\Vert x_{opt}-z_{\mathrm{opt}}\right\Vert $,
then the CDF and PDF of $R_{\mathrm{opt}}$ are derived as 
\begin{equation}
F_{R_{\mathrm{opt}}}\left(r_{\mathrm{opt}}\right)=\left(1-\mathrm{e}^{-\lambda^{\text{N\_MS}}\pi r_{\mathrm{opt}}^{2}}\right)^{\bar{N}{}_{T_{y}}},
\end{equation}
\begin{align}
f_{R_{\mathrm{opt}}}\left(r_{\mathrm{opt}}\right) & =2\pi\bar{N}_{T_{y}}\lambda^{\text{N\_MS}}r_{\mathrm{opt}}\cdot\mathrm{e}^{-\lambda^{\text{N\_MS}}\pi r_{\mathrm{opt}}^{2}}\nonumber \\
 & \cdot\left(1-\mathrm{e}^{-\lambda^{\text{N\_MS}}\pi r_{\mathrm{opt}}^{2}}\right)^{\bar{N}_{T_{y}}-1},
\end{align}
respectively.

And then the expectation of the interference received at MS $z_{\text{opt}}$
from BS $x_{\text{opt}}$ can be similarly derived as 
\begin{align}
 & \mathbb{E}\left(I_{xz_{\mathrm{opt}}}\right)=\int_{0}^{\infty}f_{R_{\mathrm{opt}}}\left(r_{\mathrm{opt}}\right)\cdot\sum_{k=1}^{K}\bar{N}_{k}P_{k}\cdot\,l\left(r_{\mathrm{opt}}\right)\,\mathrm{d}r_{\mathrm{opt}}\nonumber \\
 & =\frac{\sum_{k=1}^{K}\lambda_{k}P_{k}^{\frac{2}{\alpha}+1}}{\sum_{k=1}^{K}\lambda_{k}P_{k}^{\frac{2}{\alpha}}}\cdot\int_{0}^{\infty}f_{R_{\mathrm{opt}}}\left(r_{\mathrm{opt}}\right)\cdot\,l\left(r_{\mathrm{opt}}\right)\,\mathrm{d}r_{\mathrm{opt}}.\label{eq:expection_I_xz_multi_tier_opt}
\end{align}

\section{Numerical results\label{sec:Numerical-Results}}

In this section, we present the analytical results of the proposed
BS association scheme and compare it to the conventional scheme that
is subject to severe co-channel interference. To avoid the singularity
of path loss function $l\left(\cdot\right)$ in (\ref{eq:expection_I_xz_multi_tier})
and (\ref{eq:expection_I_xz_multi_tier_opt}), we use $l\left(r\right)=\left(1+r^{\alpha}\right)^{-1}$
\cite{ganti2012tcom} instead of $l\left(r\right)=r^{-\alpha}$ in
deriving the analytical results.

\begin{figure}
\begin{centering}
\includegraphics[width=0.9\columnwidth]{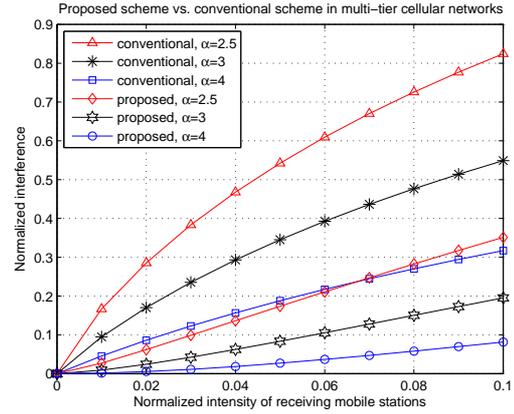}
\par\end{centering}

\caption{Interference in a 3-tier cellular network where $P_{k}=\left\{ 10,1,0.1\right\} $,
$\lambda_{k}^{\text{BS}}=\left\{ 0.01,0.1,1\right\} $ and $\lambda^{\text{BS}}/\lambda^{\text{N\_BS}}=3$.\label{fig:Interference-in-3-tier}}
\end{figure}

\begin{figure}
\begin{centering}
\includegraphics[width=0.9\columnwidth]{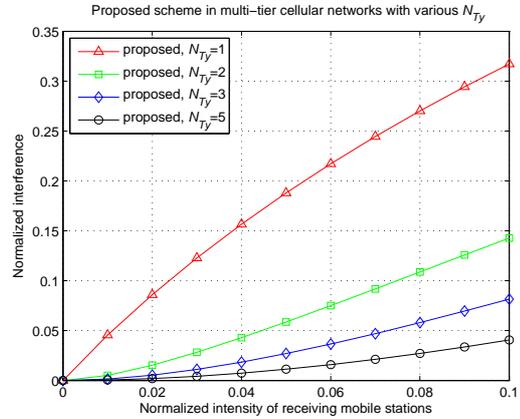}
\par\end{centering}

\caption{Interference in a 3-tier cellular network where $P_{k}=\left\{ 10,1,0.1\right\} $,
$\lambda_{k}^{\text{BS}}=\left\{ 0.01,0.1.1\right\} $ and $\alpha=4$.\label{fig:Interference-in-3-tier_Nty}}
\end{figure}

In Fig. \ref{fig:Interference-in-3-tier}, the interference in a 3-tier
cellular network is demonstrated and compared between the proposed
scheme and the conventional scheme. The BS transmission powers in
tier 1, 2 and 3 are $10$, $1$ and $0.1$ respectively, and the intensities
of BSs in tier 1, 2 and 3 are $0.01$, $0.1$ and $1$ respectively.
The result indicates that the proposed scheme is effective to reduce
the interference in multi-tier cellular networks. Fig. \ref{fig:Interference-in-3-tier}
also shows that for both the proposed interference minimized user
association scheme and the conventional scheme, with more severe path
loss effect, the interference caused to other co-channel MSs is reduced.

In Fig. \ref{fig:Interference-in-3-tier_Nty}, we show how interference
is affected by various values of $N_{T_{y}}$. A greater $N_{T_{y}}$
means that there are more candidate BSs to chose from such that it
is more probably to select a BS which leads to less interference to
other co-channel receiving MSs. Whereas when $N_{T_{y}}\rightarrow1$,
the proposed scheme degenerates to the conventional scheme.

\section{Conclusions\label{sec:Conclusions}}

In this paper, we first propose a transmission power normalization
analysis model, which significantly simplifies the analysis of the
received signal and interference, thus SIR, in multi-tier HCNs. Then
we propose an interference minimized user association scheme, which
can be applied in both single-tier and multi-tier HCNs. Using the
proposed TPNM, we proceed to analyze the interference in multi-tier
HCNs. Results demonstrate that the proposed scheme significantly reduces
the down-link interference in multi-tier HCNs, meanwhile the constraint
on the QoS of users is satisfied.

\begin{acknowledgements}
The authors would like to acknowledge the support from the International
Science \& Technology Cooperation Program of China (Grant No. 2014DFA11640,
0903 and 2012DFG12250), the National Natural Science Foundation of
China (NSFC) (Grant No. 61471180, 61271224 and 61301128), NSFC Major
International Joint Research Project (Grant No. 61210002), the Hubei
Provincial Science and Technology Department (Grant No. 2013CFB188
and 2013BHE005), the Fundamental Research Funds for the Central Universities
(Grant No. 2013ZZGH009, 2013QN136, 2014TS100 and 2014QN155), the Special
Research Fund for the Doctoral Program of Higher Education (Grant
No. 20130142120044), and EU FP7-PEOPLE-IRSES (Contract/Grant No. 247083,
318992 and 610524). This research is also supported by Australian
Research Council Discovery projects DP110100538 and DP120102030.
\end{acknowledgements}

\bibliographystyle{spbasic}
\bibliography{TPNS_paper_ref}

\end{document}